\documentclass[a4paper,twoside]{article}

\newcommand{\affil}[1]{$^{\rm #1}$}
\usepackage[authoryear]{natbib}
\bibpunct{(}{)}{;}{a}{}{,}
\usepackage{graphicx}
\date{} 
\usepackage{graphicx}

\title{\large\bf\flushleft Multivariate characterization of hydrogen Balmer emission 
  in cataclysmic variables}
\author{\parbox{\textwidth}{\flushleft
\vspace{-0.5cm}
{\it Gordon E. Sarty\affil{A,C} and Kinwah Wu\affil{B}}\\
\vspace{0.4cm}
{\small \affil{A}\,Departments of Psychology and Physics \& Engineering Physics, University of Saskatchewan, 9 Campus Drive, 
Saskatoon, Saskatchewan S7N 5A5, Canada}\\
{\small \affil{B}\,Mullard Space Science Laboratory, University College London, 
  Holmbury St.~Mary, Surrey RH5 6NT, United Kingdom and TIARA, Department of Physics, 
    National Tsing Hua University, Hsinchu 300, Taiwan  }\\
{\small \affil{C}\,Email: gordon.sarty@usask.ca}}}
\begin{document}
\twocolumn[
\begin{changemargin}{.8cm}{.5cm}
\begin{minipage}{.9\textwidth}
\vspace{-1cm}
\maketitle
%
%
\small{\bf Abstract:} 
The ratios of hydrogen Balmer emission line intensities 
     in cataclysmic variables are signatures of the physical processes that produce them. 
To quantify those signatures relative to classifications of cataclysmic variable types, 
     we applied the multivariate statistical analysis methods 
     of principal components analysis and discriminant function analysis 
     to the spectroscopic emission data set of \citet{williams_gw83}. 
The two analysis methods reveal two different sources of variation 
     in the ratios of the emission lines. 
The source of variation seen in the principal components analysis 
     was shown to be correlated with the binary orbital period. 
The source of variation seen in the discriminant function analysis 
     was shown to be correlated with the equivalent width of the H$\beta$ line.
Comparison of the data scatterplot with scatterplots of theoretical models shows 
     that Balmer line emission from T CrB systems is consistent 
     with the photoionization of a surrounding nebula. 
Otherwise, models that we considered 
     do not reproduce the wide range of Balmer decrements, 
     including ``inverted'' decrements, seen in the data. 


\medskip
\medskip
\end{minipage}
\end{changemargin}
]
\small

\section{Introduction}

Cataclysmic variables (CVs) are semi-detached binaries 
  in which a red (K or M) main-sequence, or slightly evolved, low-mass star
loses mass to a white-dwarf primary through overfilling its Roche lobe. 
They can be divided into two groups, the magnetic CVs (mCVs) and the non-magnetic CVs,  
  according to whether the white-dwarf has a magnetic field 
  that is strong enough to affect the accretion dynamics.   
Semi-detached binaries frequently possess an accretion disc,  
  where gravitational energy is converted to kinetic energy and then to radiation. 
With mCVs the magnetic-field strength of the white-dwarf is high enough to disrupt 
  the accretion disc at the magnetosphere.     
The white-dwarf's magnetosphere will determine the inner radius of the accretion disc and   
  the tidal force exerted by the secondary star will truncate the outer region of the disc.
  If the magnetosphere is large enough, no disc will form.
  
MCVs without an accretion disc are known as polars (or AM Herculis systems). 
In a polar the white-dwarf magnetic field is sufficiently strong ($\sim 10$~MG)
   to also lock the white-dwarf and its companion star  
   into synchronous (or almost synchronous)  rotation with the binary orbit.   
The accretion flow near the white-dwarf is channelled by the field and 
a shock is often formed near the base of the the accretion column 
   before the accreting gas settles onto the white-dwarf atmosphere. 
The shock heats the accreting gas, 
   leading to the emission of bremsstrahlung X-rays 
   that can subsequently photoionize the cooler gas upstream.   
For systems consisting of a white-dwarf with a moderately strong magnetic field ($\sim 1$~MG), 
   the white-dwarf spin and the orbital motion are not synchronous, 
   although the secondary may still be tidally locked and rotate with the orbit. 
These moderately magnetic systems are known as intermediate polars (IPs).  
Most IPs possess an accretion disc, 
   but  the inner region of the disc is truncated by the magnetic field, 
   and the flow is channelled in the white-dwarf magnetosphere, 
   similar to polars. 
A shock can be formed in the accretion column, 
   and the X-rays from the post-shock region can
   photoionize both the pre-shock gases in the accretion column and 
   the cool gas in the accretion disc.  
 It is worth noting that some IPs may not have a fully developed accretion disc.  
They probably have a thin accreting annulus, 
  which couples loosely to the magnetosphere of the white-dwarf.
 
Non-magnetic CVs form a heterogeneous class of objects, 
   which include nova, dwarf nova and nova-like systems, 
   primarily on the basis of their light curve behaviour 
   (although strict classification is sometimes not possible, e.g. some novae are also polars). 
They are asynchronously rotating systems,  
   and they have a fully developed accretion disc 
   with an inner edge extending to the white-dwarf surface.  
A substantial fraction of the accretion energy is dissipated 
   in a transition layer between disc and the white-dwarf. 
The strong UV photons emitted from the transition region 
   can be a main cause of photoionization in the accreting gas.  
   
CVs  are strong emission line objects, particularly in quiescence. 
They all show a clear H Balmer emission sequences. 
He I, He II and high ionization lines of other species   
   are also prominent in the optical spectra of many systems.  
The high ionization lines of the CVs are believed 
   to be emission from photoionized gas
   in the accretion disc (for IPs and non-magnetic CVs), 
   the accretion stream, or irradiatively heated surface of the companion star. 
Some of the lines may originate from hot spots in the system, 
   for example in regions where the accretion stream impacts onto the accretion disc 
   and where the accretion flow begins to couple with the magnetic field. 
 In these cases the gas is often heated by shocks 
   that are formed as consequence of an abrupt change in the dynamics of the accretion flow. 
(For reviews of CVs and related accretion physics, 
  see \citet{warner}, \citet{cropper_90}, \citet {wu_00}, \citet{wu_03}.)
    
The lines from photoionized gases and shock heat\-ed gases have different properties. 
An example is that of the relative strengths of the H Balmer emission lines. 
For instance, simple photoionization-re\-com\-bin\-a\-tion models \citep{osterbrock}
   generally predict ratios of $\approx 3.5$ for H$\alpha$/H$\beta$; 
   yet observations show H$\alpha$/H$\beta$ 
   which deviate significantly from these model predictions.  
Such deviation implies that 
   the line formation process is more complex than 
   simple photoionization and subsequent recombination.   
  
Here we analyse the H Balmer emission from CVs systematically 
  using multivariate statistical methods. 
We consider the systems in \cite{williams_gw83} as the sample for our analysis.    
We search for correlations between the emission line ratios and CV classification 
   and review the line formation mechanisms that may be responsible for the correlations. 
The paper is organized as follows. 
In \S \ref{S2} we review the multivariate statistical methods 
   used to analyze the H Balmer emission data. 
These methods are discriminant function and principal component analysis. 
In \S \ref{S3} we present the data in a four dimensional scatterplot space 
  and give the results of the discriminant function and principal component analysis 
  in that geometric space. 
We also review models of Balmer emission relevant to CVs. 
In section \S \ref{S4} we discuss possible physical causes, 
  based on existing models, of the statistical behaviour of the data. 
Two significant directions of variation are found that correlate with
the binary orbital period and the equivalent width of the h$\beta$ line. 
Our conclusion is given in \S \ref{S5}. 

\hspace*{2cm}

\section{Statistical Methods}\label{S2}

We consider the systems listed in \citet{williams_gw83} as our sample. 
The spectra were obtained in 1980 or 1981 
   using the 1.3 m telescope at McGraw-Hill Observatory. 
As the data were collected with the same telescope/instruments 
  and were reduced in a similar manner, 
  this ensures that any variations present are less likely to be caused 
  by different experimental settings and data analysis procedures.    
The emission line ratios 
  H$\alpha$/H$\beta$, H$\gamma$/H$\beta$, H$\delta$/H$\beta$ and H$\epsilon$/H$\beta$  
  were derived from the spectroscopic data presented in  \citet{williams_gw83} and
represent integrated line intensities above the continuum.  
Only systems for which all four H Balmer emission lines were measured were retained and 
   sources since identified as non-CVs or were not listed in \citet{downes} 
  (TU Leo, Z And, HK Sco and CL Sco) were discarded. 
The remaining data (representing 95 of 153 original spectra) 
    are reproduced in Table \ref{table1} 
    along with four classification schemes. None of the systems considered here
where in outburst when the data were taken. There are three observations of
two AM CVn stars which have very little hydrogen. However, the AM CVn data
were retained because their H Balmer line ratios were reported. We have also not
given multiple observations of a single star any special status in our analysis assuming,
as a first approximation, that the data represent a random sample from the observable
CVs.

Of the classification schemes considered, the first classification is that from the Atlas of CVs: 
  The Living Edition \citep{downes} as of February, 2005. 
Based on the Downes et al. classification, 
  three other, simpler, classification schemes were produced. 
For classification scheme Class 1, 
  the sources were grouped into one of four groups: 
  (i) dwarf novae,  (ii) polar \& IP (i.e.\ mCVs),
  (iii) nova \& nova-like, and (iv) double degenerate.  
(Double degenerate systems contain two white-dwarfs instead of a white-dwarf and a red dwarf.) 
All sources classified as DQ Herculis systems were put in the polar \& IP (mCV) group 
  unless the DQ Her classification was listed as very uncertain. The results of a discriminant
  function analysis (see below) of the groups in Class 1 lead to the definition of the groups
  in Class 2. That is, the groups of Class 2 were determined by a statistical analysis
  of the physically distinct groups of Class 1.
For classification scheme Class 2 
   the sources were split into two groups, (i) dwarf nova and (ii) other, 
   based on the grouping of Class 1. 
For classification scheme Class 3 
  the sources were split into two groups, (i) dwarf nova and (ii) mCVs, 
  based on the grouping of Class 1, 
  which excluded the nova \& nova-like and the double degenerate groups. The grouping of
  Class 3 was defined in an attempt to find statistical differences between objects that are
  similar but for the magnetic field of the primary.

\begin{table*}
\centering
\begin{minipage}{140mm}
\caption{Balmer line emission data as taken from \citet{williams_gw83} together with the
classifications used in the discriminant function analysis. \label{table1}}

\ 
 
\hspace*{-1em}
\footnotesize
\begin{tabular}{@{}cccccccccccc@{}}
\hline
Star$^{a}$	&	\underline{H$\alpha$}	&	\underline{H$\gamma$}	&	\underline{H$\delta$}	&	\underline{H$\epsilon$}	&	ACV$^{b}$ 	&	
\multicolumn{3}{c}{\ \ \ \ \ \ Class$^{c,d,e}$}	& EW($\beta$)$^{f}$
& DW($\beta$)$^{g}$ & $P_{\mbox{orb}}$$^{h}$ \\
 & H$\beta$ & H$\beta$ & H$\beta$ & H$\beta$ &   Class & 1 & 2 & 3 & (\AA) & (km/s) & (days) \\ 
\hline
V368 Aql	&	0.98	&	0.92	&	0.87	&	0.85	&	na	    &	N\&NL &		O &	    &     	& 		& 0.3452	\\
V603 Aql	&	0.49	&	1.33	&	1.50	&	1.70	&	na/dq::	&	N\&NL &		O &	  	& 7.7 	& 		& 0.1385	\\
V603 Aql	&	0.44	&	1.22	&	1.43	&	1.55	&	na/dq::	&	N\&NL &		O &	  	& 6.2 	& 		& 0.1385	\\
DN Gem		&	0.53	&	1.17	&	1.43	&	1.58	&	na		&	N\&NL &		O &	  	& 8.4 	& 562 	& 0.12785	\\
DQ Her		&	0.79	&	1.13	&	1.35	&	1.30	&	na/dq	&	P\&IP &		O &	M 	& 33.3 	& 505 	& 0.193621	\\
DQ Her		&	0.59	&	1.09	&	1.26	&	1.30	&	na/dq	&	P\&IP &		O &	M 	& 16.7 	& 519 	& 0.193621	\\
DQ Her		&	0.59	&	1.13	&	1.21	&	1.26	&	na/dq	&	P\&IP &		O &	M 	& 20.5 	& 460 	& 0.193621	\\
V533 Her	&	0.43	&	1.29	&	1.62	&	1.64	&	na		&	N\&NL &		O &	  	& 4.9 	& 396 	& 0.147	\\
DI Lac		&	0.58	&	1.13	&	1.30	&	1.22	&	na		&	N\&NL &		O &	  	& 		& 		& 0.543773	\\
HR Lyr		&	0.52	&	1.25	&	1.37	&	1.43	&	na		&	N\&NL &		O &	  	& 		&		& 	\\
BT Mon		&	0.75	&	1.04	&	1.08	&	1.08	&	na		&	N\&NL &		O &	  	& 23.3 	& 343 	& 0.338814	\\
BT Mon		&	0.92	&	0.99	&	0.96	&	0.93	&	na		&	N\&NL &		O &	  	& 35.0 	& 592 	& 0.338814	\\
BT Mon		&	0.87	&	1.04	&	1.08	&	1.09	&	na		&	N\&NL &		O &	  	& 30.6 	& 644 	& 0.338814	\\
V841 Oph	&	0.78	&	1.05	&	1.17	&	1.13	&	nb		&	N\&NL &		O &	  	& 1.6 	& 		& 0.6014	\\
GK Per		&	1.25	&	0.71	&	0.65	&	0.62	&	na/dq	&	P\&IP &		O &	M  	& 10.8 	& 309 	& 1.9968	\\
X Ser		&	0.56	&	1.06	&	1.37	&	1.17	&	nb:		&	N\&NL &		O &	  	& 7.1 	& 		& 1.48	\\
T CrB		&	2.97	&	0.65	&	0.46	&	0.29	&	nra		&	N\&NL &		O &	  	& 		& 		& 227.5687	\\
RS Oph		&	3.37	&	0.60	&	0.43	&	0.43	&	nra		&	N\&NL &		O &	  	& 12.7	& 		& 455.72 \\
T Pyx		&	0.56	&	1.24	&	1.41	&	1.50	&	nrb		&	N\&NL &		O &	  	& 9.7 	& 554 	& 0.076223	\\
V1017 Sgr	&	1.42	&	0.47	&	0.50	&	0.34	&	nb		&	N\&NL &		O &	  	& 		& 		& 5.714	\\
RX And		&	0.84	&	1.16	&	1.21	&	1.30	&	ugz		&	DN	&	DN	&	DN	& 58.3 	& 664 	& 0.209893 \\
RX And		&	0.54	&	1.52	&	1.77	&	1.79	&	ugz		&	DN	&	DN	&	DN	& 		& 		& 0.209893 \\
AR And		&	0.97	&	1.26	&	1.29	&	1.63	&	ug		&	DN	&	DN	&	DN	& 38.5 	& 702 	& 0.163 \\
AR And		&	0.84	&	1.16	&	1.63	&	1.71	&	ug		&	DN	&	DN	&	DN	& 39.2 	& 379 	& 0.163 \\
UU Aql		&	0.97	&	1.05	&	1.20	&	1.24	&	ug		&	DN	&	DN	&	DN	& 50.4 	& 454 	& 0.163532 \\
UU Aql		&	0.97	&	1.16	&	1.39	&	1.55	&	ug		&	DN	&	DN	&	DN	& 80.9 	& 670 	& 0.163532 \\
SS Aur		&	1.06	&	1.20	&	1.27	&	1.50	&	ug		&	DN	&	DN	&	DN 	& 		& 		& 0.1828	\\
SS Aur		&	1.13	&	1.08	&	1.16	&	1.37	&	ug		&	DN	&	DN	&	DN 	& 122.0 & 569 	& 0.1828	\\
FS Aur		&	0.83	&	1.12	&	1.25	&	1.45	&	ug		&	DN	&	DN	&	DN	& 60.8 	& 903 	& 0.0595\\
FS Aur		&	0.86	&	1.06	&	1.14	&	1.38	&	ug		&	DN	&	DN	&	DN	& 35.3 	& 571 	& 0.0595 \\
Z Cam		&	0.48	&	1.17	&	1.36	&	1.56	&	ugz		&	DN	&	DN	&	DN 	& 		& 		& 0.289841	\\
HT Cas		&	0.75	&	1.16	&	1.26	&	1.64	&	ugsu	&	DN	&	DN	&	DN 	& 97.7 	& 943 	& 0.073647	\\
HT Cas		&	0.78	&	1.13	&	1.38	&	1.72	&	ugsu	&	DN	&	DN	&	DN	& 117.0 & 1032 	& 0.073647 \\
WW Cet		&	0.96	&	1.22	&	1.46	&	1.70	&	ugz:	&	DN	&	DN	&	DN 	& 42.7 	& 849 	& 0.1758	\\
WW Cet		&	0.64	&	1.13	&	1.22	&	1.27	&	ugz:	&	DN	&	DN	&	DN 	& 25.4 	& 673 	& 0.1758	\\
1E0643.0	&	0.85	&	1.08	&	1.33	&	1.59	&	ug/ugz	&	DN	&	DN	&	DN	& 96.5 	& 776 	& 0.216778 \\
1E0643.0	&	1.18	&	0.91	&	0.82	&	1.14	&	ug/ugz	&	DN	&	DN	&	DN	& 81.3 	& 584 	& 0.216778 \\
1E0643.0	&	1.02	&	0.72	&	0.78	&	0.86	&	ug/ugz	&	DN	&	DN	&	DN	& 43.5 	& 461 	& 0.216778 \\
1E0643.0	&	0.99	&	0.96	&	1.02	&	1.22	&	ug/ugz	&	DN	&	DN	&	DN 	& 64.0 	& 506 	& 0.216778	\\
1E0643.0	&	0.53	&	1.25	&	1.47	&	1.67	&	ug/ugz	&	DN	&	DN	&	DN 	& 6.1 	& 		& 0.216778	\\
1E0643.0	&	0.81	&	1.08	&	1.13	&	1.33	&	ug/ugz	&	DN	&	DN	&	DN	& 49.1 	& 873 	& 0.216778 \\
SY Cnc		&	0.73	&	0.99	&	1.09	&	1.09	&	ugz		&	DN	&	DN	&	DN 	& 13.7 	& 498	& 0.38 \\
SY Cnc		&	0.54	&	1.42	&	1.56	&	1.64	&	ugz		&	DN	&	DN	&	DN	& 		& 		& 0.38 \\
YZ Cnc		&	0.72	&	1.34	&	1.45	&	1.70	&	ugsu	&	DN	&	DN	&	DN	& 80.1 	& 941 	& 0.0868 \\
YZ Cnc		&	0.66	&	1.33	&	1.58	&	2.06	&	ugsu	&	DN	&	DN	&	DN	& 131.0 & 906 	& 0.0868\\
AK Cnc		&	0.72	&	1.00	&	0.91	&	1.01	&	ugsu	&	DN	&	DN	&	DN	& 		& 		& 0.0651 \\
SS Cyg		&	1.13	&	0.89	&	0.92	&	1.05	&	ugss	&	DN	&	DN	&	DN	& 69.5 	& 477 	& 0.27513 \\
SS Cyg		&	1.05	&	0.91	&	0.86	&	0.97	&	ugss	&	DN	&	DN	&	DN 	& 67.0 	& 513	& 0.27513 \\
SS Cyg		&	0.70	&	1.18	&	1.29	&	1.46	&	ugss	&	DN	&	DN	&	DN	& 30.2 	& 786 	& 0.27513 \\
EM Cyg		&	0.60	&	1.10	&	1.33	&	1.42	&	ugz		&	DN	&	DN	&	DN	& 2.6 	& 		& 0.290909\\
AB Dra		&	0.59	&	1.16	&	1.35	&	1.50	&	ugz		&	DN	&	DN	&	DN	& 19.6 	& 973 	& 0.152 \\
IR Gem		&	0.77	&	1.16	&	1.37	&	1.60	&	ugsu	&	DN	&	DN	&	DN 	& 80.6 	& 554	& 0.0684 \\
AH Her		&	0.97	&	0.96	&	0.98	&	1.06	&	ugz		&	DN	&	DN	&	DN 	& 26.5 	& 682 	& 0.258116	\\
EX Hya		&	0.49	&	1.16	&	1.34	&	1.59	&	ug/dq	&	P\&IP	&	O	&	M & 54.4 & 939 	& 0.068234	\\
EX Hya		&	0.55	&	1.12	&	1.25	&	1.38	&	ug/dq	&	P\&IP	&	O	&	M & 75.2 & 1492	& 0.68234\\
EX Hya		&	0.61	&	1.18	&	1.30	&	1.39	&	ug/dq	&	P\&IP	&	O	&	M & 80.7 & 1144 & 0.68234	\\
\end{tabular}
\normalsize
\end{minipage}
\end{table*}

\setcounter{table}{0}
\begin{table*}
\centering
\begin{minipage}{140mm}
\caption{{\it --- Continued.}}

\ 

\hspace*{-2em}
\footnotesize
\begin{tabular}{@{}cccccccccccc@{}}
\hline
Star\footnote{G61$-$29 = GP Com, YY Dra = DO Dra, 2A0311$-$227 = EF Eri, N Sgr 1962 = V3890 Sgr,
2H2215$-$086 = FO Aqr, 2H2252$-$035 = AO Psc, 0633+71 = BZ Cam}
&	\underline{H$\alpha$}	&	\underline{H$\gamma$}\	&	\underline{H$\delta$}\	&	\underline{H$\epsilon$}\	&	
ACV\footnote{Atlas of Cataclysmic Variables (ACV) \citep{downes} nomenclature: ug = U Gem variable (dwarf nova), ugz = U Gem variable (Z Cam subtype), ugss = U Gem variable (SS Cyg subtype),
ugsu = U Gem variable (SU UMa subtype), ugwz = U Gem variable (WZ Sge subtype),
na = Fast nova, nb = Slow nova, nra = Recurrent nova - giant donor, nrb = Recurrent nova - non-giant donor,
nlv = Novalike variable (V Sge subtype), ux = Novalike variable (UX UMa subtype), vy = Novalike variable (VY Scl subtype - systems which undergo low states), am = AM Herculis variable (synchronous rotators), dq = DQ Herculis variable (non-synchronous rotators), ibwd = Interacting binary white-dwarf, : = Uncertain, :: = Very uncertain}
&	\multicolumn{3}{c}{\ \ \ \ \ \ Class\footnote{DN = dwarf nova, P\&IP = polar and IP,
N\&NL = nova and NL, DD = double degenerate}\footnote{DN = dwarf nova, O = other}\footnote{DN =
dwarf nova, M = magnetic}} & EW($\beta$)\footnote{EW = equivalent width}  
& DW($\beta$)\footnote{DW = Doppler width} & $P_{\mbox{orb}}$\footnote{Orbital period data from \citet{downes}}	\\
 &  H$\beta$	& 	H$\beta$	&	H$\beta$	&	H$\beta$  &   Class & 1  & 2  & 3 & (\AA) & (km/s) & (days) \\ 
\hline
T Leo	&	0.60 	&	1.42 	&	1.50	&	1.54	&	ugsu	&	DN	&	DN	&	DN & 115.0 & 457 & 0.05882	\\
X Leo	&	1.05	&	1.21 	&	1.38	&	1.67	&	ug	&	DN	&	DN	&	DN	& 22.3 & 919 & 0.1644 \\
CN Ori	&	0.49	&	1.33	&	1.54	&	1.59	&	ugz	&	DN	&	DN	&	DN	& & & 0.163199 \\
CZ Ori	&	1.05	&	1.10	&	1.13	&	1.37	&	ug	&	DN	&	DN	&	DN	& 43.5 & 503 & 0.2189 \\
RU Peg	&	0.60	&	1.21	&	2.81	&	3.35	&	ugss	&	DN	&	DN	&	DN	& 4.7 & & 0.3746 \\
WZ Sge	&	0.47	&	1.30	&	1.55	&	1.71	&	ugwz/dq	&	P\&IP	&	O	&	M & 27.8 & 725	& 0.056688 \\
SW UMa	&	0.42	&	1.35	&	1.51	&	1.58	&	ugsu/dq	&	P\&IP	&	O	&	M	& & & 0.056815 \\
SW UMa	&	0.66	&	1.22	&	1.41	&	1.63	&	ugsu/dq	&	P\&IP	&	O	&	M & 64.3 & 657 & 0.056815	\\
SW UMa	&	0.68	&	1.18	&	1.39	&	1.55	&	ugsu/dq	&	P\&IP	&	O	&	M & 71.6 & 825 & 0.056815	\\
TW Vir	&	0.79	&	1.21	&	1.41	&	1.50	&	ug	&	DN	&	DN	&	DN & 69.8 & 805 & 0.18267	\\
TW Vir	&	0.91	&	1.26	&	1.38	&	1.99	&	ug	&	DN	&	DN	&	DN	& 96.5 & 601 & 0.18267 \\
TW Vir	&	0.91	&	1.21	&	1.35	&	1.75	&	ug	&	DN	&	DN	&	DN	& 73.3 & 680 & 0.18267 \\
AE Aqr	&	1.30	&	0.71	&	0.61	&	0.52	&	dq	&	P\&IP	&	O	&	M & 21.7 & 369 & 0.411656	\\
TT Ari	&	0.44	&	1.30	&	1.47	&	1.57	&	vy/dq:	&	P\&IP	&	O	&	M	& 7.0 & & 0.13755 \\
G61$-$29	&	0.64	&	1.21	&	1.42	&	1.50	&	ibwd	&	DD	&	O	&   & & & 0.032339	\\
AM CVn	&	0.37	&	1.43	&	1.71	&	1.83	&	ibwd	&	DD	&	O	&   & & & 0.011907		\\
AM CVn	&	0.37	&	1.43	&	1.75	&	1.93	&	ibwd	&	DD	&	O	&  	& & & 0.011907	\\
YY Dra	&	1.41	&	1.08	&	1.18	&	1.29	&	dq	&	P\&IP	&	O	&	M	& 129.0 & 739 & 0.165374 \\
2A0311$-$227	&	1.09	&	0.96	&	0.91	&	0.97	&	am	&	P\&IP	&	O	&	M & 43.9 & 901 & 0.056266	\\
AM Her	&	0.98	&	1.43	&	1.59	&	1.84	&	am	&	P\&IP	&	O	&	M & 18.4 & 230	& 0.128927 \\
AM Her	&	0.85	&	1.27	&	1.39	&	1.55	&	am	&	P\&IP	&	O	&	M & 15.2 & 232	& 0.128927 \\
AM Her	&	0.72	&	1.26	&	1.46	&	1.59	&	am	&	P\&IP	&	O	&	M	& 55.4 & 517  & 0.128927 \\
AM Her	&	1.01	&	1.16	&	1.34	&	1.47	&	am	&	P\&IP	&	O	&	M	& 48.8 & 355 & 0.128927 \\
V426 Oph	&	0.66	&	1.18	&	1.34	&	1.45	&	ugz/dq:	&	P\&IP	&	O	&	M & 14.9 & 813 & 0.2853	\\
V442 Oph	&	0.54	&	1.10	&	1.22	&	1.25	&	vy	&	N\&NL	&	O	&  	& 4.8 & 469	& 0.12433 \\
V2051 Oph	&	0.60	&	0.98	&	1.18	&	1.35	&	ugsu	&	DN	&	DN	&	DN	& 124.0 & 853 & 0.062428\\
VV Pup	&	0.62	&	1.25	&	1.24	&	1.20	&	am	&	P\&IP	&	O	&	M	& 93.9 & 513 & \\
V Sge	&	0.50	&	1.25	&	1.59	&	1.71	&	nlv	&	N\&NL	&	O	&  	& 52.9	& 739 & 0.514197 \\
V Sge	&	0.61	&	1.24	&	1.43	&	1.62	&	nlv	&	N\&NL	&	O	&  	& 62.2 & 645 & 0.514197 \\
N Sgr 1962	&	1.96	&	0.85	&	0.74	&	0.72	&	nra	&	N\&NL	&	O	&  	& 59.6 & 208 & 	\\
RW Tri	&	0.59	&	1.17	&	1.26	&	1.34	&	ux	&	N\&NL	&	O	&   & & & 0.231883		\\
UX UMa	&	0.47	&	1.31	&	1.51	&	1.79	&	ux	&	N\&NL	&	O	&   & 4.1 & 360	& 0.196671 \\
AN UMa	&	0.52	&	1.14	&	1.22	&	1.24	&	am	&	P\&IP	&	O	&	M & 28.1 & 783 & 0.79753	\\
AN UMa	&	0.51	&	1.14	&	1.19	&	1.36	&	am	&	P\&IP	&	O	&	M & 24.2 & 661 & 0.79753	\\
2H2215$-$086	&	0.58	&	1.21	&	1.42	&	1.42	&	dq	&	P\&IP	&	O	&	M & 25.0 & 382 & 0.20206	\\
2H2215$-$086	&	0.47	&	1.21	&	1.33	&	1.34	&	dq	&	P\&IP	&	O	&	M & 27.3 & 323	& 0.20206 \\
2H2215$-$086	&	0.52	&	1.21	&	1.43	&	1.42	&	dq	&	P\&IP	&	O	&	M & 17.6 & 704 & 0.20206	\\
2H2252$-$035	&	0.47	&	1.26	&	1.46	&	1.51	&	dq	&	P\&IP	&	O	&	M & 10.6 & 582 & 0.149626	\\
0623+71	&	0.43	&	1.37	&	1.53	&	1.64	&	vy	&	N\&NL	&	O	&   & & & 0.153693		\\
\hline
\end{tabular}
\normalsize
\end{minipage}
\end{table*}

\begin{figure*}
\vspace*{6.5in}
\includegraphics{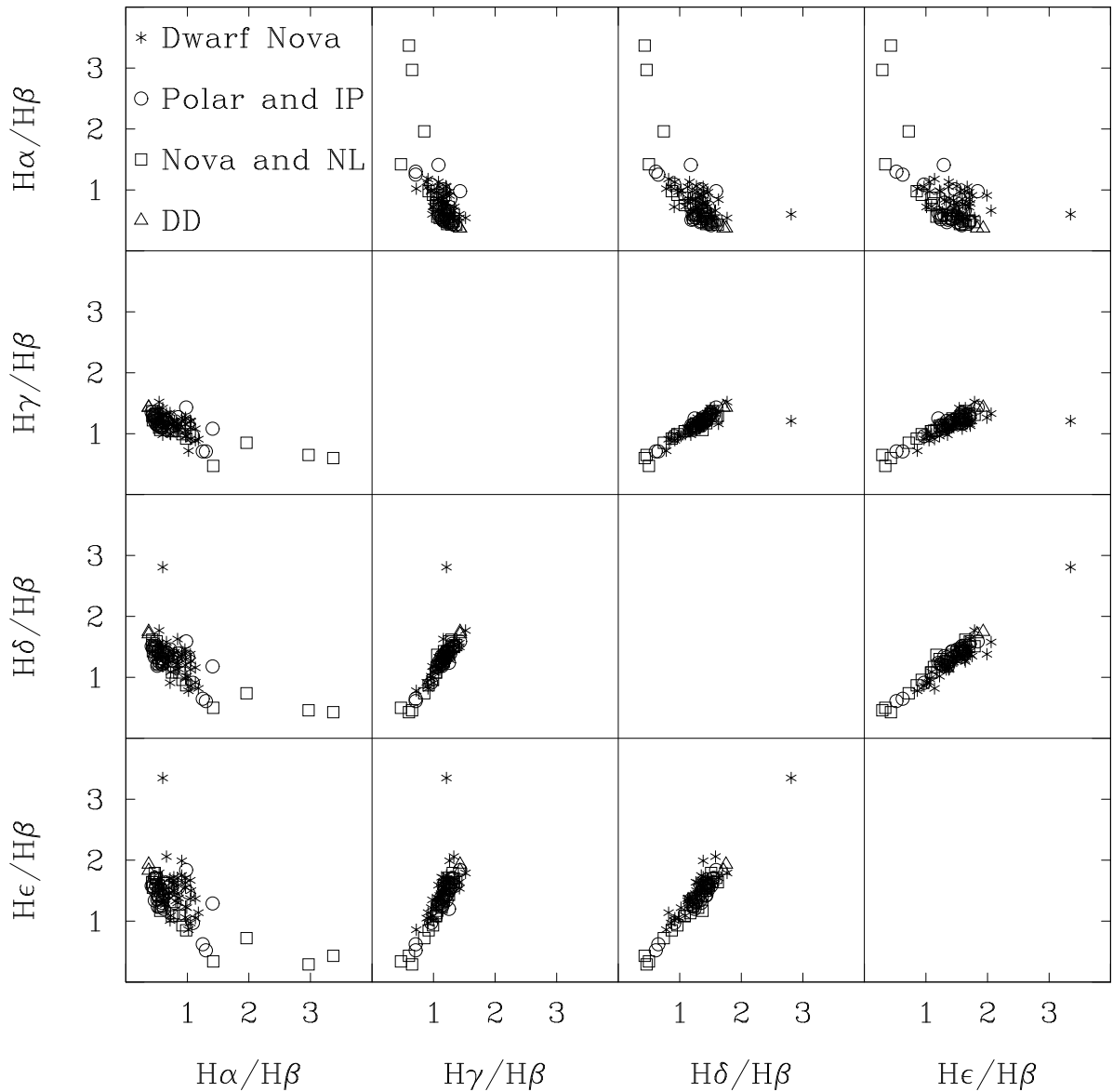}
\caption{
Scatterplot of the emission line data given in Table \ref{table1}. 
IP = intermediate polar, NL = nova-like, DD = double degenerate 
  (which are systems containing two white-dwarfs). 
Individual systems are difficult to identify in general at this scale 
  but four outlying nova and nova-likes and one outlying dwarf nova are obvious. 
From top to bottom in the H$\gamma/$H$\beta$ versus H$\alpha/$H$\beta$ box, 
  the nova and/or nova-likes are RS Oph, T CrB, N Sgr 1962 and V1017 Sgr. 
The outlying dwarf nova is RU Peg. 
See text for a discussion of those stars. 
\label{fig1}}
\end{figure*}

Discriminant functions were constructed, using the SPSS statistical software 
(SPSS Inc., Chicago, USA), 
to maximally separate the two groups of Class 2 and Class 3. 
The discriminant functions were of the form
\begin{equation}\label{df}
   z = c_{0} + \sum_{i=1}^{4} c_{i} x_{i}
\end{equation}
where $x_{i}$ are the emission line ratios and $c_{i}$ are coefficients 
such that the values of $z$ for the centroids (mean) of the two groups are maximally 
different ($c_{0}$ is arbitrary). Specifically we used $x_{1}$ to represent H$\alpha/$H$\beta$, 
$x_{2}$ to represent H$\gamma/$H$\beta$,
$x_{3}$ to represent H$\delta/$H$\beta$ 
and $x_{4}$ to represent H$\epsilon/$H$\beta$. Discriminant functions may be used to quantify the separation of
more than two groups; we originally computed a discriminant function for the classification scheme of Class 1 but found
significant separations only for the groups subsequently defined in Class 2 and 3.
The statistical significance of the discriminant function was characterized 
with Wilk's $\Lambda$, a statistic that generalizes Fisher's analysis of variance $F$ 
statistic to the multivariate situation, and an associated $p$-value 
that gives the probability of incorrectly rejecting the null hypothesis 
of no difference between the two groups. The null hypothesis is generally rejected
and statistically significant differences are said to be found when $p < 0.05$.
The discriminant function test of two group separation on the basis of the multivariate line ratio data 
is mathematically equivalent, under the assumption of normal homoscedasticity (which means that all groups are assumed to be multivariate
normally distributed with the same covariance matrices), to
Hotelling's $T^{2}$ test. The $T^{2}$ statistic tests if the mean vectors of two groups are
significantly different and is equivalent to $\Lambda$ because $T^{2}$ and $\Lambda$ give the same likelihood
ratio test with two groups \citep{rencher}. 
Histograms, along with the best fit normal distribution, 
   of the discriminant function values were plotted to show the group separation 
   provided by the discriminant function.
To aid in the interpretation of the resulting discriminant functions 
   a principal component analysis (PCA) was also conducted. 
   
PCA finds the best description of the data in scatterplot space assuming they are
a sample from a multivariate normal population \citep{rencher}. Surfaces of
constant $\sigma$ (standard deviation from the mean vector) of a multivariate
normal distribution form ellipsoids in scatterplot space. So PCA finds an
ellipsoid in scatterplot space such that the major axis of the ellipsoid lies in the
direction of maximum variation of the data; this is the first principal component
direction. Subsequent principal component directions are directions of maximal
variation that are geometrically orthogonal to the previous principal component directions. The ellipsoid is the geometrical expression of the data's covariance
matrix -- the eigenvalues and eigenvectors of the covariance matrix define
the size and direction of the ellipsoid. It should be noted that PCA considers all the data as coming from a single
multivariate normally distributed population with one mean vector while discriminant functions assume that
each group comes from it's own population with different mean vectors (but the covariance
matrices of each population are assumed equal). The discriminant function's
level surfaces will generally contain the principal axes of the individual group ellipsoids (since they are the same, except translated,
under the assumption of homoscedasticity) to
achieve maximal group separation. From this it can be seen that the normal to the
discriminant function's
level surfaces should be roughly perpendicular to the PCA's first component
direction defined from the ellipsoid that considers all the data together.

\begin{figure}
\vspace{3in}
\includegraphics{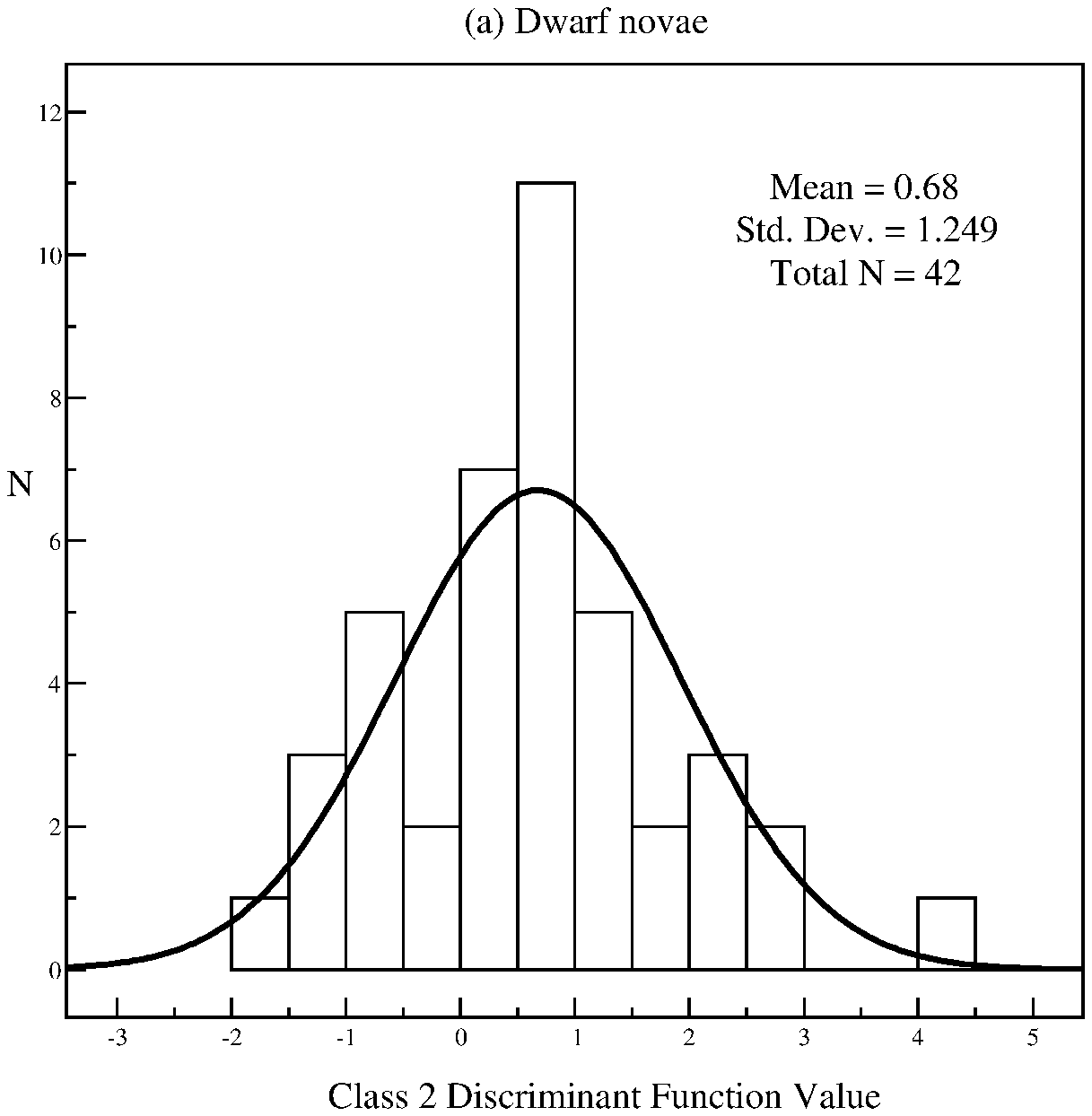}
\vspace{3in}
\includegraphics{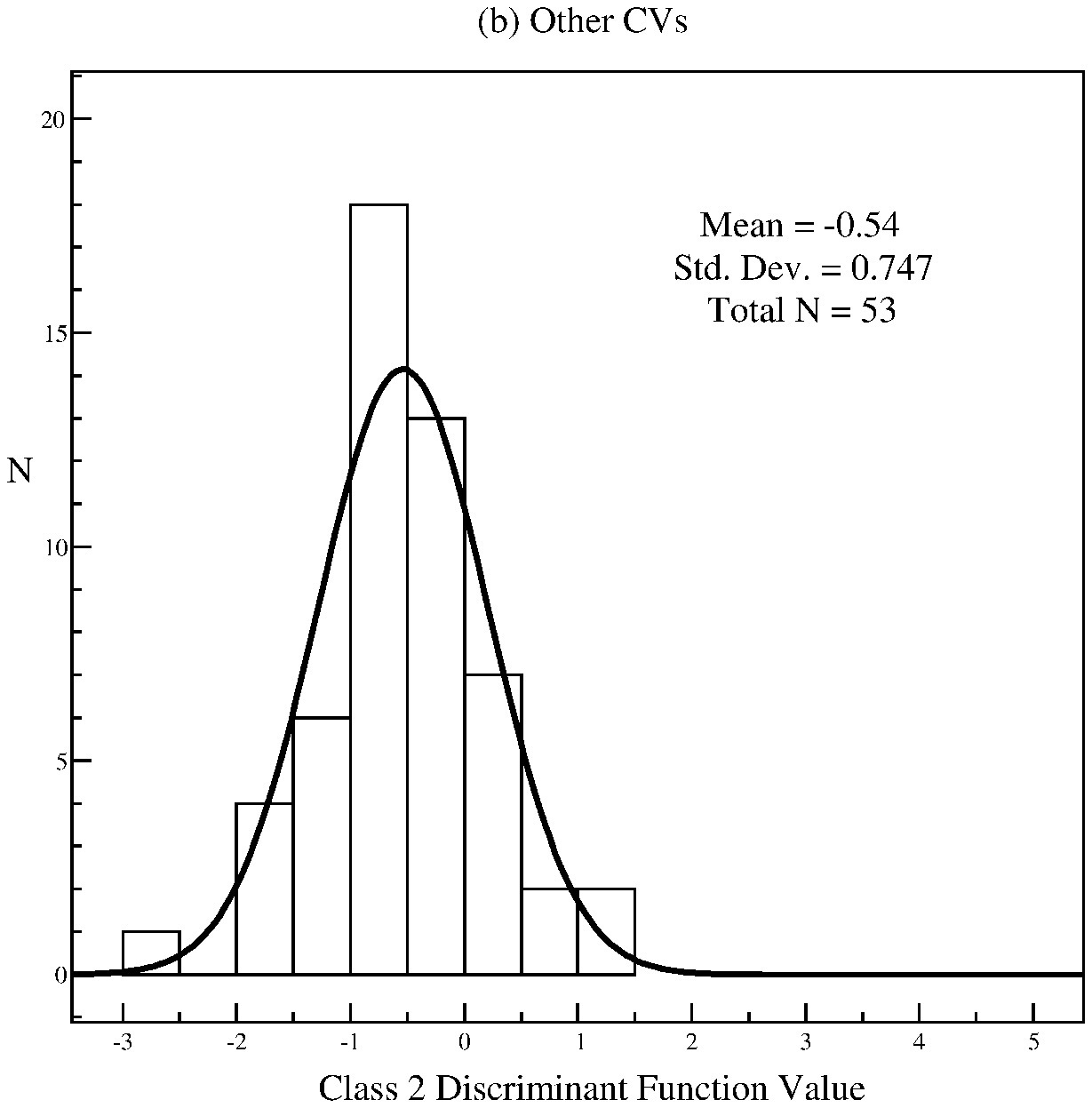}
\caption{
Histograms of discriminant function values 
  for discriminating dwarf novae from all other CVs. 
All other CVs include nova in low states and mCVs. 
(a) Dwarf novae. 
(b) Other CVs including polars, IPs, nova remnants and nova-like variables. 
Overlain on each histogram is a normal curve having the same mean and standard deviation 
  as the histogram. 
$N$ = number of systems. 
\label{fig2}}
\end{figure}

\begin{figure}
\vspace{3.0in}
\includegraphics{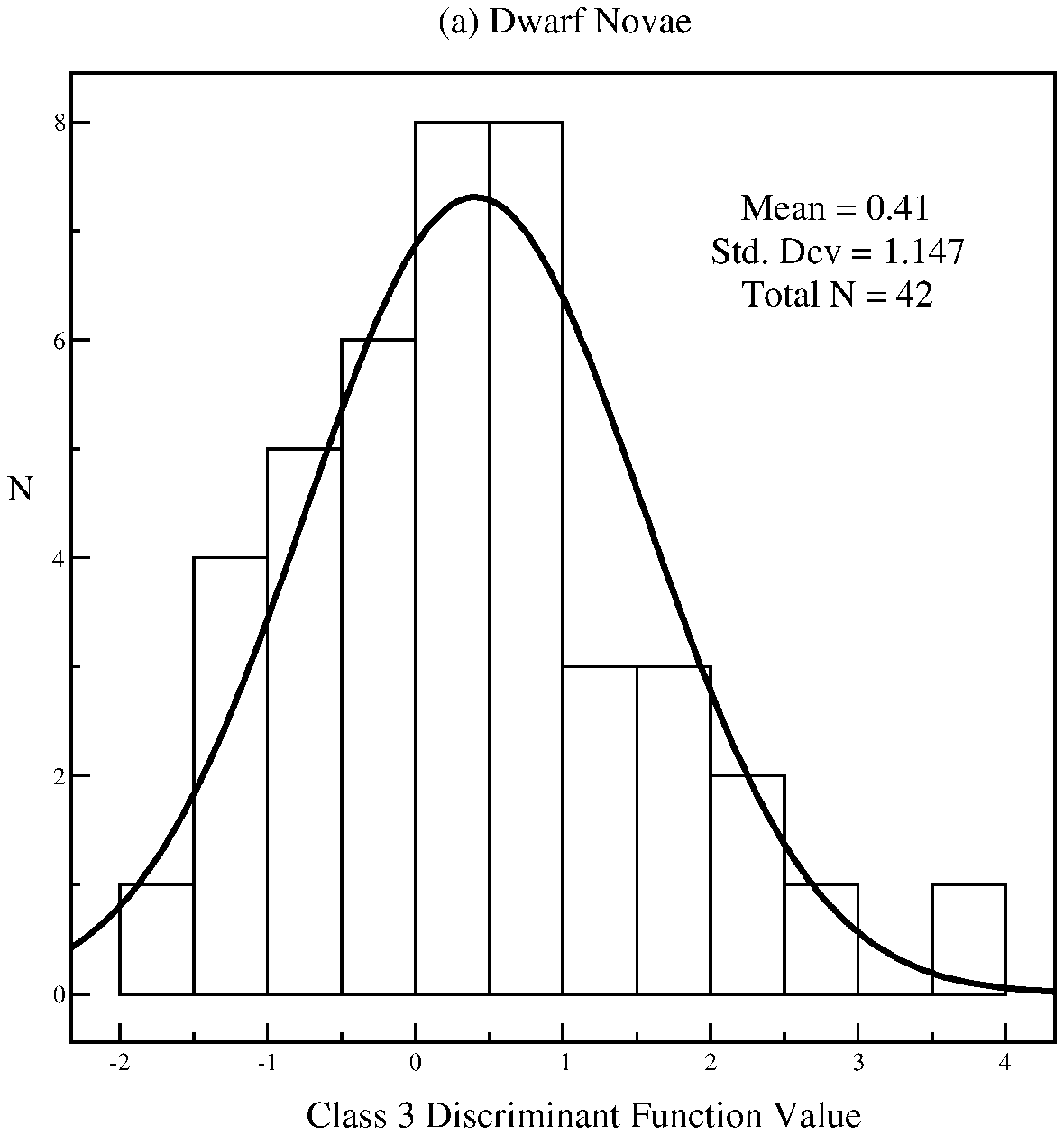}
\vspace{3.0in}
\includegraphics{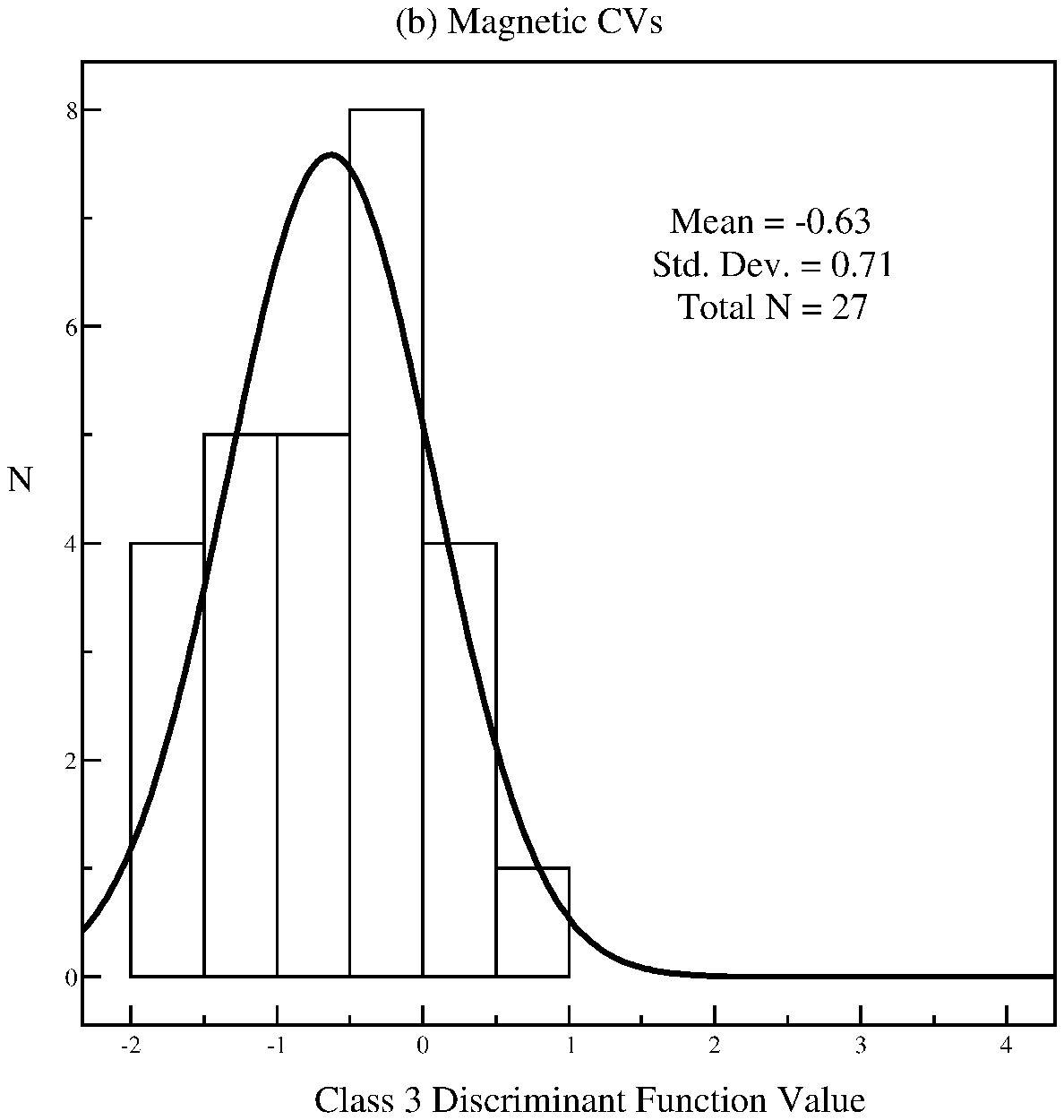}
\caption{
Histograms of discriminant function values for discriminating dwarf novae from mCVs. 
(a) Dwarf novae. 
(b) MCVs including polars and IPs. 
Overlain on each histogram is a normal curve having the same mean and standard deviation 
  as the histogram.
$N$ = number of systems. 
\label{fig3}}
\end{figure}

For comparison to the data, scatterplots of H Bal\-mer emission line ratios  
   from accretion discs predicted by theoretical models were constructed.   
Four models were considered. 
The first is the optically thin nebula model \citep{osterbrock}, 
   which assumes that the line formation is dominated by photoionization-recombination.  
The second model considered 
   the temperature and density distribution in the accretion disc explicitly 
   via an $\alpha$-disc prescription \citep{shakura}. 
The line calculations were carried out 
  assuming a local thermodynamic equilibrium (LTE) condition with emission
primarily from the outer region of the disc, which is optically thin to the continuum 
  emission \citep{williams_re80}. 
The third model assumed that the emission region is isothermal 
  and employed a non-LTE radiative transfer formulation \citep{williams_ga91} 
  in the line calculations. 
Finally we considered the chromospheric models of \citet{williams_ga95} where a
hot, photoionized chromosphere above the accretion disc is modelled.

\section{Analysis}\label{S3}

\subsection{Observed properties of H lines in the CV samples}

The scatterplot of the emission line ratio data from Table \ref{table1} is given in Fig.~\ref{fig1}. 
It may be discerned that the data tend to fall on a line in the 4D scatterplot space. 
Visible in the scatterplot are some obviously outlying systems. 
Those systems are as follows. 
RS Oph is a recurrent nova with past outbursts observed in 1898, 1933, 1958, 1967, 1985 and 2006, 
T CrB is a recurrent nova with past outbursts observed in 1866 and 1946, 
  and N Sgr 1962 (V3890 Sgr) is a  recurrent nova with past outbursts 
  observed in 1962 and 1990. 
RS Oph, T CrB and V3890 Sgr are all of the T CrB subclass of recurrent nova 
  which have M giant secondaries, 
  $P_{\rm{orb}}$   \raisebox{-.45ex}{$\stackrel{>}{\sim}$} 100d 
  and a high rate of mass accretion, 
  ${\dot M}_1$ \raisebox{-.45ex}{$\stackrel{>}{\sim}$} $1 \times 10^{-8}$~M$_{\odot}~{\rm yr}^{-1}$, 
  onto a white-dwarf primary 
  whose mass is close to the Chandrasekhar mass \citep{warner}. 
There are no other T CrB type systems in the data set.
\citet{duerbeck} lists V1017 Sgr as a possible symbiotic star or recurrent nova. 
RU Peg is a U Gem type of dwarf nova with no other known characteristics that
would suggest that it is different from other dwarf novae. 

The discriminant function that maximally split the Class 2 groups of dwarf novae versus others was
\begin{equation}
z = 1.799 -0.152 x_{1} -1.927 x_{2} -9.714 x_{3} + 9.226 x_{4}.
\end{equation}
The value of Wilk's lambda 
  for this function was $\Lambda = 0.729$ with $p = 9 \times 10^{-6}$. 
The unit vector ${\bf V_2}$ = [~$-$0.011  $-$0.142  $-$0.718  0.681~]$^{T}$ is normal 
  to the level surface hyperplanes of the discriminant function for Class 2. 
Histograms of the distribution of the values for the Class 2 discriminant function 
  are given in Fig.~\ref{fig2}.

The discriminant function that maximally split 
  the Class 3 groups of dwarf novae versus mCVs was
\begin{equation}
z = -0.583 +1.125 x_{1} -1.229 x_{2} -7.812 x_{3} + 7.757 x_{4}.
\end{equation}
The value of Wilk's lambda for this function was $\Lambda = 0.791$ with $p = 0.004$.
The unit vector ${\bf V_3}$ = [~0.101  $-$0.110  $-$0.702  0.697~]$^{T}$ is normal 
  to the level surface hyperplanes of the discriminant function for Class 3.
Histograms of the distribution of the values for the Class 3 discriminant function 
  are given in Fig.~\ref{fig3}. 
The angle between ${\bf V_2}$ and ${\bf V_3}$ is 6.8$^{\circ}$.

The direction first principal component (direction of maximum data variance) is
    given by the unit vector
   ${\bf V_1}$ = [~$-$0.441  0.510  0.530  0.514~]$^{T}$. 
The direction of a least squares line fit through the data in the 4D scatterplot space 
   is given by the unit vector
   ${\bf V_4}$ = [~$-$0.816  0.236  0.339  0.405~]$^{T}$ 
   and the angle between ${\bf V_1}$ and ${\bf V_4}$ is 29.8$^{\circ}$. 
The angles between the first principal component direction, ${\bf V_1}$, 
  and the two discriminant function directions, ${\bf V_2}$ and ${\bf V_3}$, 
  are 95.6$^{\circ}$ and 96.6$^{\circ}$ respectively.

\begin{figure*}
\vspace*{6.0in}
\includegraphics{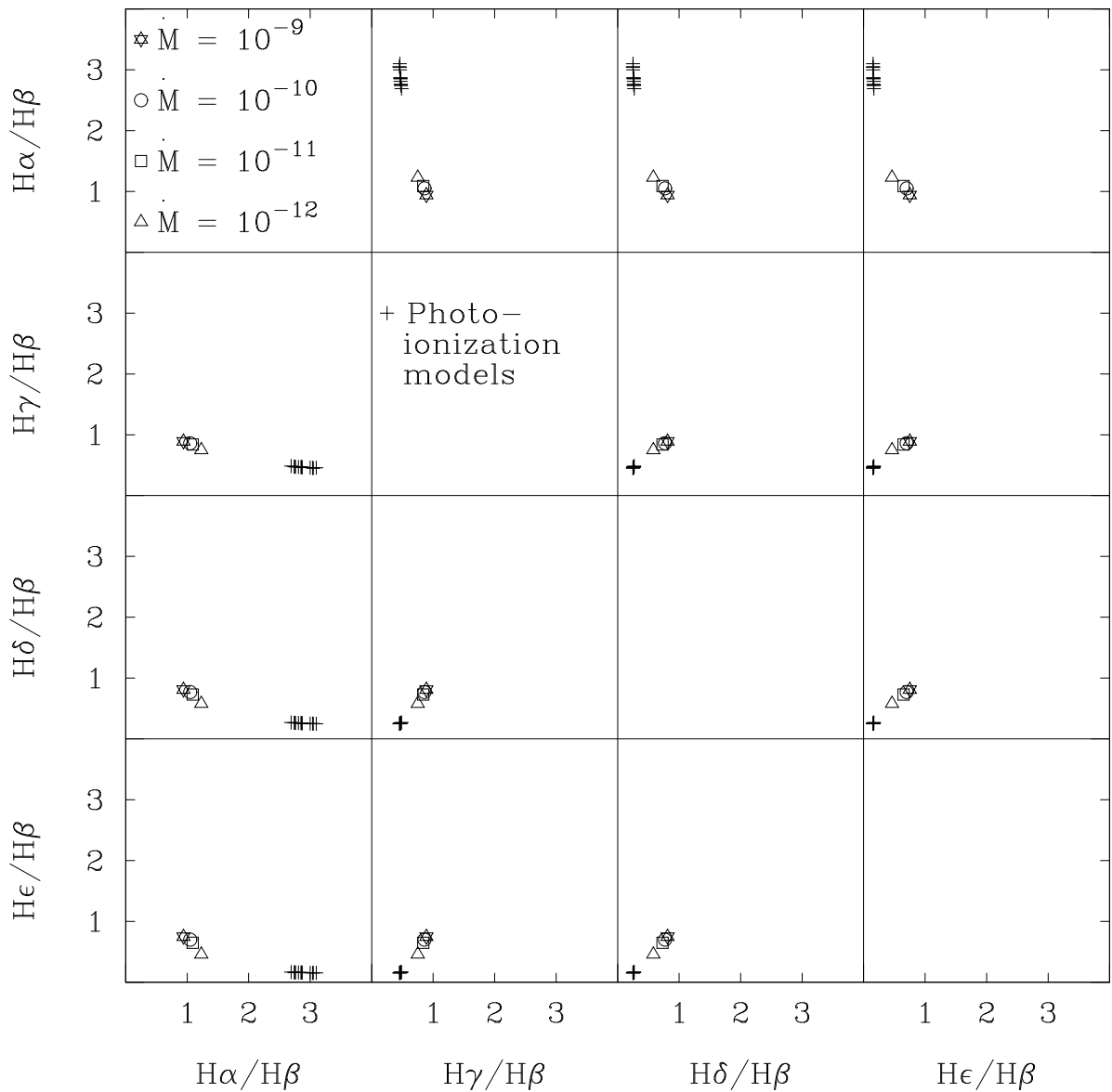}
\caption{Photoionization and LTE optically thin accretion disc models. 
The photoionization emission models are from \citet{osterbrock} 
and both Case A and Case B models (see text) are plotted.
The Balmer emission line ratios
from the LTE radiative transfer calculations assume an $\alpha$-disc model 
with various mass transfer rates \citep{williams_re80}. $\dot{M}$ is given in
M$_{\odot}$ yr$^{-1}$.
\label{fig4}}
\end{figure*}
  
\begin{figure*}
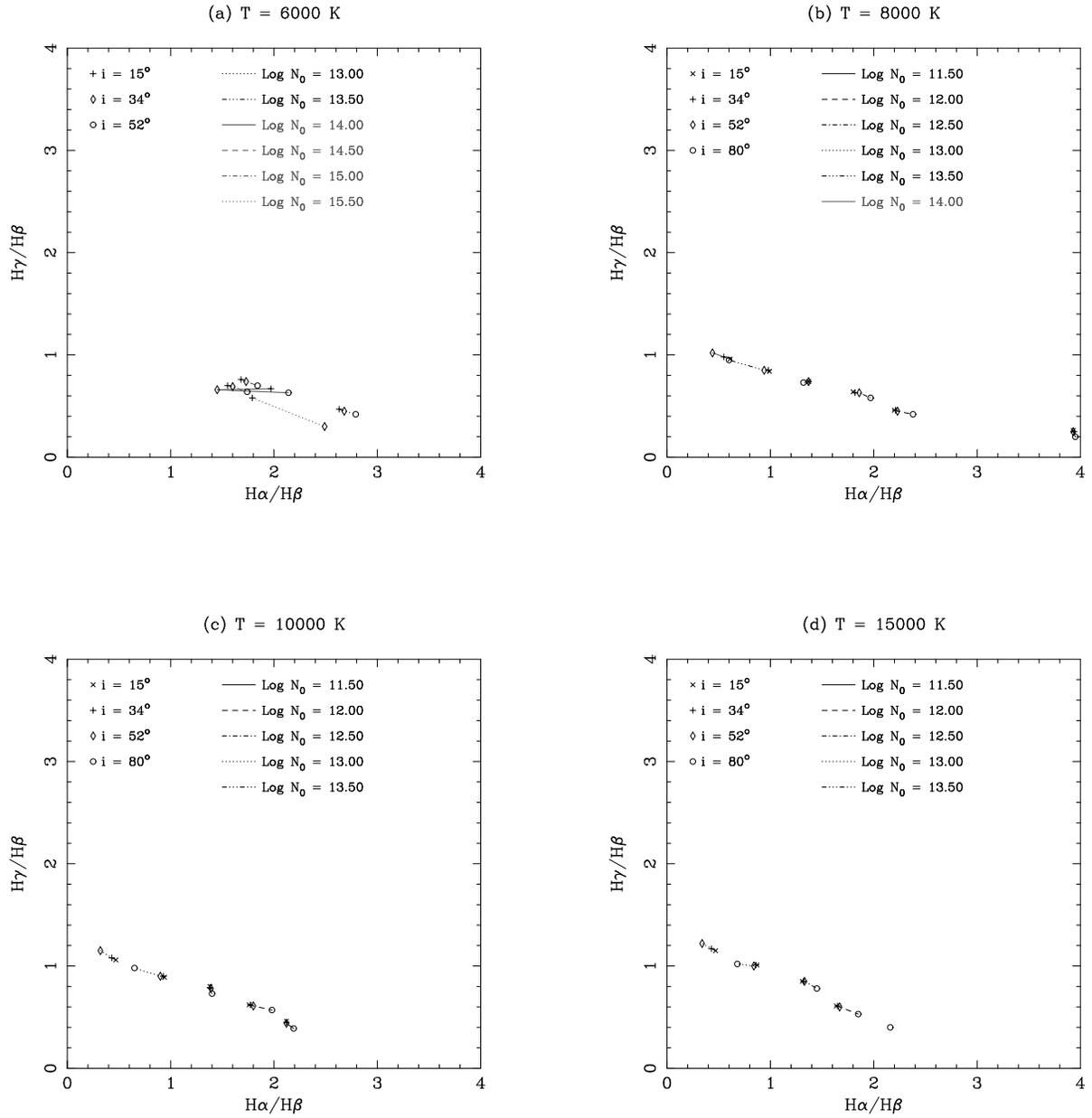

\vspace*{7.0in}
\includegraphics{Figure5a.ps}
\includegraphics{Figure5c.ps}
\includegraphics{Figure5b.ps}
\includegraphics{Figure5d.ps}
\caption{
Scatterplots of H${\alpha}$/H${\beta}$ versus H${\gamma}$/H${\beta}$ values 
  from the non-LTE calculations of \citet{williams_ga91}. 
Hydrogen number density, $N_{0}$, units are cm$^{-3}$; 
   $i$ is disc inclination. 
\label{fig5}}
\end{figure*}

\subsection{Line characteristics of theoretical models}

\subsubsection{Optically thin photoionized plasmas}

It is useful to consider emission from photoionized gas, 
  which may be present in a disc, accretion stream or in surrounding nebulousity 
left over from previous nova eruptions since it is expected that all CVs experience
periodic nova eruptions \citep{warner}.  
For this purpose the photoionization-recombination models in \citet{osterbrock} are relevant 
  and the resulting scatterplot points are given in Fig.~\ref{fig4}. 
After excitation via boundary layer X-ray and UV illumination, 
  optical recombination line emission may be modelled
  under two situations known as Case A, 
  in which all emitted photons escape from the gas, 
  and Case B, where the optical depth is high enough to convert every Lyman-line photon into
lower-series photons plus either a Lyman $\alpha$ photon or two continuum photons through scattering.
For Case B the assumed electron number density, $N_{\rm e}$ is relevant, 
  and the models illustrated in Fig.~\ref{fig4} 
  assume $N_{\rm e} = 10^{2}$, $10^{4}$ and $10^{6}$~cm$^{-3}$.
It may be seen that the dependence of the photoionization caused Balmer line ratios 
  on temperature and gas density is relatively weak 
  but that significantly non-flat Balmer decrements are predicted. 
In addition, the trend of scatterplot positions for higher density gas versus lower density gas 
   for the photoionization models is in the same general direction as the two disc models described below with,
   for example, the higher density cases being positioned in the upper left hand part of the scatterplot 
   and the lower density discs in the lower right hand part 
   for the H${\alpha}$/H${\beta}$  versus H${\gamma}$/H${\beta}$ 
(abscissa versus ordinate) scatterplot.

\subsubsection{Accretion discs with optically thin continuum}

\citet{williams_re80} presented calculations of H Balmer lines 
from accretion discs, with an outer region optically thick to the lines 
but optically thin to the continuum.  
The emission lines arise directly from a hot disc 
instead of from a disc chromospheric surface layer heated by X-ray and UV radiation 
from the accretion boundary layer near the white-dwarf surface. 
An $\alpha$-disc prescription is used to determine the radial density and temperature profile,  
and the line-centre intensities are given by the local Planck function 
at the relevant frequencies. 
\citet{williams_re80} calculations predict that 
the position of a line in the emission line ratio scatterplot space 
depends on the disc mass flow rate $\dot{M}$ (Fig.~\ref{fig4}). The predicted equivalent width
of the H$\beta$ line, EW($\beta$), varies from 0.2 \AA\ for $\dot{M} = 10^{-9}$ M$_{\odot}$
yr$^{-1}$ to 65 \AA\ for $\dot{M} = 10^{-12}$ M$_{\odot}$
yr$^{-1}$.
The cases represented in Fig.~\ref{fig4}  
are for a disc of outer radius $r_{\rm d} = 4 \times 10^{10}$~cm 
and a white-dwarf of radius $R_1 = 6 \times 10^{8}$~cm 
and mass $M_1 = 1.0~{\rm M}_{\odot}$. 
The assumed disc radius $r_{\rm d}$ corresponds to 1/3 the separation 
of two 1 M$_{\odot}$ stars having an orbital period  of 4 hr.   

We note that in order to have the assumption of an optically thin continuum 
satisfied, the accretion rate of the system must not greatly exceed 
$10^{-10}~{\rm M}_{\odot}{\rm yr}^{-1}$.  
Most CVs are, however, observed to have higher accretion rates, 
so that models assuming an optically thin continuum are not always applicable.  
 
\subsubsection{Non-LTE emission from isothermal medium}\label{nonlte}

The predictions of a non-LTE model as computed by \citet{williams_ga91} 
are summarized in Fig.~\ref{fig5}.  
The model assumes an isothermal horizontally infinite disc \citep{williams_shipman} 
  with a Gaussian dropoff in density with vertical height from the midplane. 
The H number density $N_{0}$ (in cm$^{-3}$) represents the value in the disc midplane, 
  which is generally a monotonic function of disc mass flow rate. 
The position of the line ratios on the scatterplot are more influenced 
   by the value of $N_{0}$ than by the temperature 
   with the higher density discs being represented in the upper left hand part of the scatterplot
   and the lower density discs being represented in the lower right hand position. 
This trend is similar to the LTE models in Fig.~\ref{fig4} 
  where systems with high accretion rates are positioned 
  in the upper left hand part of the H${\alpha}/$H${\beta}$ versus H${\gamma}/$H${\beta}$
(abscissa versus ordinate) box and the lower mass flow rates are represented in the lower right hand location. 
The range of predicted line ratios are, however, wider for this model 
   especially for H${\alpha}/$H${\beta}$.
In particular, the model predicts significantly non-flat Balmer decrements. Additionally,
for disc models with temperatures between 15,000 K and 8000 K, the predicted EW($\beta$)
is much larger than for the optically thin LTE models. The predicted EW($\beta$)
varies strongly as a function of $N_{0}$, going from $\sim 200$ \AA\ for $\log N_{0} = 11.5$
to $\sim 10$ \AA\ for $\log N_{0} = 14.0$.

\subsubsection{Chromospheric model}

\begin{figure}[h]
\vspace*{3in}
\includegraphics{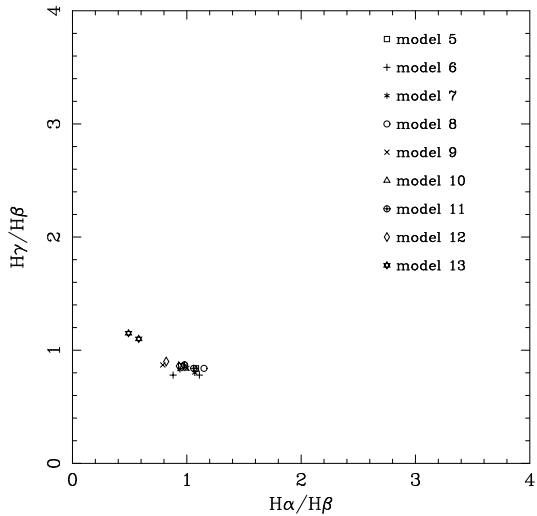}
\caption{
Chromospheric models of \citet{williams_ga95}. For each model number, from 5 to 13,
two outer disc radii are considered, $10^{10}$ and $2 \times 10^{10}$ cm with the model
position moving from the upper left to lower right as the radius is increased. The
assumed disc mass flow rate is $10^{14}$ g~s$^{-1}$ for models 5, 7 and 8, $10^{15}$ for models
6, 9 and 12, $10^{13.5}$ for models 10 and 11, and $10^{15.5}$ for model 13. The
boundary layer luminosity, which photoionizes the chromospheric gas, is taken as
compatible with the disc mass flow rate, except for models 8, 9 and 11 where the
boundary layer luminosity is that corresponding to 10 times the disc mass flow rate.
Finally the $\alpha$ disc viscosity parameter is assumed to be 1.0 for all models
except for models 7, 12 and 13 where $\alpha$ is assumed to be 0.5, 10.0 and 100.0
respectively. 
\label{fig6}}
\end{figure}

The prediction of Balmer line absolute strengths from disc chromospheres is made
difficult because of the strong dependence of the underlying continuum contributions
on the disc model \citep{ferguson97}. However models of disc chromospheres have been made because
of the probable existence of such chromospheres.
Situations where $Q/\kappa$ is a decreasing
function of temperature, where $Q$ is the energy generation and $\kappa$ is the Rosseland
mean absorption coefficient in an atmosphere, can lead to an instability and a discontinuity
in temperature in which the temperature jumps higher as the pressure decreases \citep{shaviv}.
These are situations that lead to the formation of chromospheres and/or coronas.
Such a situation was shown to exist in $\alpha$ prescription accretion disc models by
\citet{adam}. Given the probable existence of disc chromospheres, \citet{williams_ga95}
modelled emission from chromospheres over $\alpha$-disc models using radiative transfer
methods, particularly computations using the Feautrier method, similar to those used in
the non-LTE isothermal models described in \S \ref{nonlte} above. The Balmer line emission
ratios predicted by \citet{williams_ga95} chromosphere models are summarized in Fig.~\ref{fig6}.
The associated predicted EW($\beta$) are small and vary from 1.02 \AA\ in model 6 to
32.27 \AA\ in model 11.

\section{Discussion}\label{S4}

\cite{echevarria} has compiled emission line data for CVs from several sources 
  and produced a number of relevant scatterplots 
  but did not subject the data to any formal multivariate statistical analysis.
Echevarr\'{i}a reported a good fit (unquantified) of H Balmer emission line ratio scatterplot 
   with an LTE gas slab model computed by \citet{drake} 
   in which position along roughly the principal component direction 
   found in our present analysis depended only on disc temperature. 
Sources positioned further from the Drake \& Ultrich model line 
  in the H Balmer emission line ratio scatterplot space were found by Echevarr\'{i}a 
  to have narrow H$\beta$ equivalent widths.  
Here we have taken H Balmer emission line data from a single survey \citep{williams_gw83} 
   and performed PCA and discriminant function analysis. 

The two multivariate analysis methods that we used literally give orthogonal results as
we have indicated in \S \ref{S2} would be expected if the discriminant function were found to be statistically significant. 
The first principal component direction is roughly in line with a least squares line fit 
through the 4D scatterplot data; 
the direction of the least squares line, ${\bf V_{4}}$,  is 29.8$^{\circ}$ 
from the first PCA direction, ${\bf V_{1}}$. 
The discriminant function directions  ${\bf V_{2}}$ and ${\bf V_{3}}$, 
on the other hand, form angles of 95.6$^{\circ}$ and 96.6$^{\circ}$, respectively, 
with the first PCA direction. 
Further, the discriminant function directions are nearly orthogonal 
to the second and third principal component directions 
and is most nearly aligned with the fourth principal component direction 
(within $\sim 15{^\circ}$ --- 
there is no statistical or geometrical reason for the discriminant function direction 
to match any of the principal component directions, its direction is strongly
dependent on the direction of the difference between the mean vectors of the groups).
The difference between the first principal component direction and the discriminant direction 
suggests that the physical processes responsible for the majority of the variance 
of the data (82\% of the overall variance is in the first principal component direction) 
is a different one than what separates dwarf nova from other CVs. The first principal
component direction reflects the fact that the Balmer decrement changes in a systematic
way from one system to the next. We also note that the ``decrements'' become ``increments'' for many
systems (cf. Fig.~\ref{fig1} where it may be seen that many ratios are greater than
one for H$\gamma/$H$\beta$ to H$\epsilon/$H$\beta$). None of the models considered here (or
any model that we are aware of) predict such ``increments'' (cf. Figs.~\ref{fig4}--\ref{fig6}). 

\begin{figure}
\vspace*{3.5in}
\includegraphics{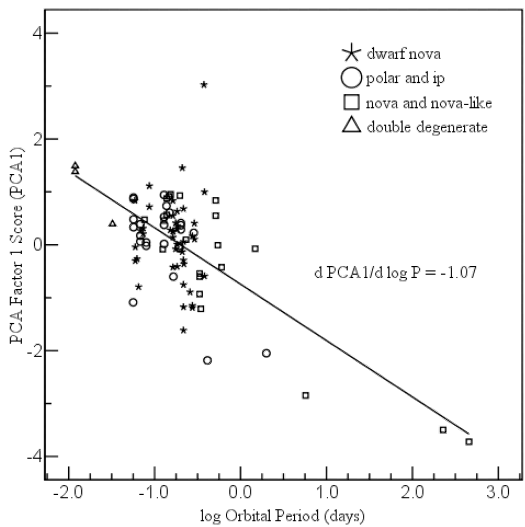}
\caption{
Correlation of the log of the orbital period, as given by \citet{downes}, 
  with the first principal component value. 
Shown is the least squares line fit through the data; a fit having a correlation of $-$0.664. 
\label{fig7}}
\end{figure}

\citet{patterson} shows a good correlation for low-mass interacting binaries 
  between the orbital period $P_{\rm orb}$ 
  and the mean mass transfer rate $\langle {\dot M} \rangle$ from the secondary to the primary, 
  as determined by a variety of methods, 
  and presents an empirical relation 
\begin{equation}
\label{mdot}
  \langle {\dot M} \rangle\  
    = \  5.1^{+3}_{-2} \times 10^{-10} P_{4}^{3.2 \pm 0.2} ~{\rm M}_\odot~{\rm yr}^{-1} \ , 
\end{equation}
   where $P_{4}$ is $P_{\rm orb}/4$ with $P_{\rm orb}$ given in days. Although mass transfer
   rates for individual systems are known to vary widely from the rate given in equation (\ref{mdot}),
we can apply it to the data set of \citet{williams_gw83} to see if any average
driven trends can
identified. 
Using $P_{\rm orb}$ as given by \citet{downes}, $\dot{M}$ was computed 
   using equation (\ref{mdot}) 
   and the correlation between $\dot{M}$ and the first principal component data values determined; 
   a correlation of $-$0.436 ($p = 1 \times 10^{-5}$, two-tailed) was found. 
More directly, the correlation between $\log P_{\rm orb}$ and
   the first principal component data values was found to be $-$0.664 
   ($p = 5 \times 10^{-13}$, two-tailed), 
   a result which may by appreciated in Fig.~\ref{fig7}. 
The implication is that the majority of the variance in the data 
  along the first principal component direction is due to variance in mass transfer rate. 
However, from inspection of Fig.~\ref{fig7}, 
   it appears that this conclusion may be strongly influenced by the T CrB systems 
   which are known to have high mass flow rates \citep{warner}. 
Removal of the T CrB systems only reduces the correlation 
   between the first principal component and $\log P_{\rm orb}$ 
   to $-$0.452  ($p = 8 \times 10^{-6}$, two-tailed) 
   so the influence of the T CrB systems did not overwhelmingly bias the correlation.

The ordinate of the graph shown in Fig.~\ref{fig7} roughly 
  follows the direction (within 29.8$^{\circ}$) of the line 
  along which the data points lie in the 4D emission line ratio scatterplot space. 
The LTE model, 
  as shown in Fig.~\ref{fig4}, predicts a mass flow rate that increases 
  as the first principal component coordinate increases. 
This variation is incompatible with $\dot{M}$ increasing with $P_{\rm orb}$, 
   as given by equation (\ref{mdot}) for example, 
   and with an increase of $P_{\rm orb}$ with a decreasing first principal component coordinate 
   as shown in Fig.~\ref{fig7}. A possible reason for the opposing correlations between
$\langle \dot{M} \rangle$ and the first principal component coordinate seen in the data and between
$\dot{M}$ and the first principal component seen in the LTE models
is that the LTE models may only be applicable to individual systems as an approximation
when $\dot{M}$ varies because of disc instability and cannot
be generalized to CV population behaviour where $\langle \dot{M} \rangle$ varies because of
orbital period differences.  

Discriminant function analyses on the basis of contrasting dwarf novae with all other CVs or of
contrasting dwarf novae with magnetic CVs both lead to similar separations.
The mean discriminant function values for dwarf novae are 
statistically significantly different from other CVs 
  and, from inspection of Figs.~\ref{fig2} and \ref{fig3}, 
  the variance of H Balmer emission line discriminant function values is wider 
  for dwarf novae than for other CVs\footnote{It should be noted that interstellar
reddening is a possible source of variance for all groups, but this should not
affect the {\em difference} in variances either between dwarf novae and other CVs or
between dwarf novae and magnetic CVs.}. 
Since the discriminant function directions are nearly orthogonal 
  to the first principal component direction, 
  the physical processes that lead to the segregation of dwarf novae from other CVs
  in the discriminant function direction appear not to be related directly 
  to the differences in mass transfer rate if equation (\ref{mdot}) is taken as true. 
Supporting this conclusion, we additionally found no significant correlation 
  between $\langle \dot{M} \rangle$, as given by equation (\ref{mdot}), and discriminant function values 
  ($p > 0.75$, two-tailed) or between $\log P_{\rm orb}$ and discriminant function values 
  (two-tailed $p > 0.75$ with the marginal exception of a correlation of $-$0.213, 
  with two-tailed $p = 0.041$, 
  between $\log P_{\rm orb}$ and the Class 2 discriminant function values 
  --- but any kind of correction for multiple comparisons makes this correlation non-significant).

Visual comparison of Fig.~\ref{fig1} to Fig.~\ref{fig4} suggests that 
  the main source of  H Balmer emission from T CrB systems is 
  from nebula-like photoionization plasmas. 
Such photoionization may occur in the nebulousity surrounding the binary star,
    left behind from the recurrent nova activity. The scatterplot values for the
    four T CrB systems represented in the dataset are, however, more widely spaced
    than the values given by the theoretical calculations. This increase in spacing may be caused
by the underlying absorption spectrum of the giant star which might tend to reduce
H$\alpha$/H$\beta$. An exception is RS Oph
which shows line ratio values consistent with a low temperature (5000 K) Case A
(optically thin) situation.

Possible sources of H Balmer emission in polars include the stream 
   between the secondary and the (magnetic) threading region, 
   gas falling down the magnetic field line,
   the heated surface of the stream close to the accretion zone 
   or a reprocessing of high energy accretion radiation on the surface 
   of the secondary star. 
Since polars possess no disc 
 and all other CVs likely do possess accretion discs of some sort, 
 one would think that difference would discriminate polars 
 from all other CVs in the 4D scatterplot space examined here. 
But that is not the case. 
Dwarf nova discs, however, may be distinguished from other CV discs 
  in that the disc is unstable,  
  and they undergo thermal cycling which leads to outbursts. 
The differences in H Balmer emission line ratios between polars and non-dwarf nova CVs 
  is smaller than the differences between dwarf novae and all other CVs, 
  at least in the direction of the discriminant functions found here. 
There therefore appear to be physical processes 
  occurring in the accretion discs of dwarf novae 
  that are different from those happening with the other CV types 
  and that there is a wide variation in the manner 
  in which those unique dwarf nova processes occur. The obvious interpretation
is that, since the state of the accretion discs in dwarf novae are continuously
changing, our data includes discs at random phases between their last outburst
and the next one.
  
\begin{figure}[h]
\vspace*{3.5in}
\includegraphics{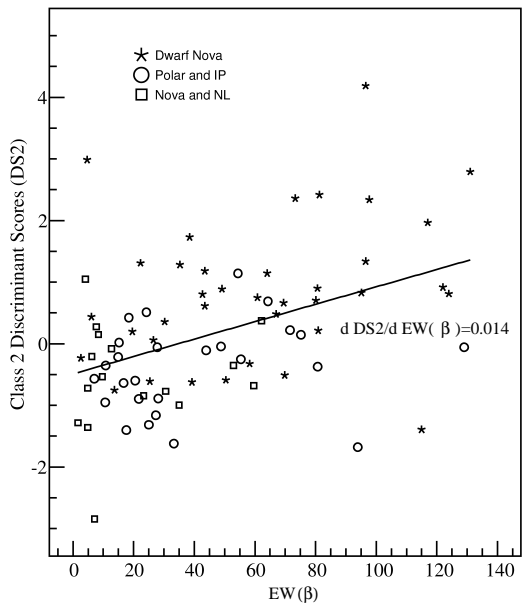}
\caption{Correlation of the equivalent width of H$\beta$, EW($\beta$), as given by 
\citet{williams_gw83}, with the value of the discriminant
function based on Class 2. Shown is the least squares line fit through the data; a fit having
a correlation of 0.422. No EW($\beta$) values for AM CVn stars were given by Williams. \label{fig8}}
\end{figure}
  
\cite{echevarria} speculated that variation normal to our principal component direction
was due to variation in the equivalent width of the H$\beta$
line, EW($\beta$).
To test the hypothesis that variation along the Class 2 discriminant function
was related to equivalent width, we computed the correlation of the Class 2 discriminant function values 
with EW($\beta$), as given by \citet{williams_gw83},
and found a correlation of $r = 0.422$ ($p = 1 \times 10^{-4}$,
two-tailed). This correlation is illustrated in Fig.~\ref{fig8}. This correlation, in turn,
implies a wide variance in EW($\beta$) of dwarf novae in the dataset, relative to other CVs, from the
wide variance of their discriminant function values (see Fig.~\ref{fig2}). The correlation
of EW($\beta$) with the first principal component value is insignificant
at $r = -0.016$ ($p = 0.891$) indicating that the cause of EW($\beta$) is
closely related to the cause of variation in the discriminant function direction.
\citet{patterson_85} show that the ratio of X-ray flux (0.2 -- 4.0 keV) to optical flux, $F_{x}/F_{v}$,
is positively correlated with EW($\beta$) for CVs with accretion disks. If this were true
for all CVs then the discriminant function direction would
correspond to the relative X-ray intensity of the accretion region on the white dwarf. However,
this interpretation is inconsistent with observed $0.001 < F_{x}/F_{v} < 10$ for dwarf novae and $3 < F_{x}/F_{v} < 300$
for polars \citep{warner}
when the values of the discriminant function as shown in Fig.~\ref{fig2} or \ref{fig3} are considered. In
other words, polars tend to have smaller EW($\beta$) than dwarf novae, not larger widths as would be
implied by a larger $F_{x}/F_{v}$. The physical processes
behind the production of Balmer emission in polars is not a continuous variation of the processes that
produce Balmer emission in dwarf novae.

The Doppler width of the H$\beta$ line, DW($\beta$), is weakly
correlated with both the Class 2 discriminant function value ($r = 0.325$,
$p = 0.006$) and the first principal component value ($r = 0.377$, 
$p = 0.001$).  Therefore factors influencing the DW($\beta$) value, such
as broadening due to disc rotation, have no coherent effect on the
variation of Balmer emission line ratios in the population of CVs. But the variation in
radiative transfer processes that give rise to variation of EW($\beta$) is
coherently related to the variation of emission line ratios in our
sample of CVs, specifically in the discriminant function direction in which
dwarf nova may be distinguished from all other types of CVs. Inspection of
Fig.~\ref{fig8} reveals that the variation of dwarf nova EW($\beta$) values
in the discriminant function direction
is much larger that the variation of other CV EW($\beta$) values, which
tend to be smaller. This behaviour, of course, is simply a reflection of the behaviour
of discriminant function values illustrated in Fig.~\ref{fig2}. Since all dwarf nova contain unstable accretion discs \citep{cannizzo} as a
primary feature distinguishing them from other CVs, we may speculate that
the variation of dwarf nova EW($\beta$) values is due to the variation of
physical processes in the disc, perhaps related to thermal cycling in the disc as mentioned earlier.

\cite{echevarria} essentially found variation along 
  what we have defined as the first principal component direction 
  to be a function of temperature while we hypothesized 
  that it is essentially a function of mass transfer rate. 
The two conclusions are consistent  
  because higher mass flow rates implies higher temperatures 
  in the $\alpha$-disc model (\citet{frank}).  
Echevarr\'{i}a found deviation in H Balmer line scatterplot space 
  away from the LTE model of Drake \& Ultrich 
  to be related to the equivalent width of the H$\beta$ line. 
We analogously find, more rigorously, that variation in the Class 2
discriminant function direction, which is roughly orthogonal 
  to the first principal component direction, 
   is correlated with the equivalent width of the H$\beta$ line.

\section{Conclusions}\label{S5}

Our analysis of hydrogen Balmer lines in CVs shows that 
    the source of variation seen in the principal components analysis 
    are correlated with the binary orbital period, 
    and the source of variation seen in the discriminant function analysis 
    are correlated with the equivalent width of the H$\beta$ line. 
Comparing models to the data  produces one positive result. 
 The ratios of hydrogen Balmer lines predicted by the photoionized model 
  match the observed line ratios of the T CrB systems. In particular, the emission
from RS Oph is consistent with photoionization from a $\sim$5000 K optically thin (Case A)
nebula.    
These systems, with a massive white dwarf and likely progenitors of type Ia supernovae,   
   are hypothesized to be embedded in a nebula 
   produced by the frequent novae outbursts \citep{warner}.  
Our analysis supports this scenario, 
   as the observed Balmer emissions could originate from a photoionized nebula. 
Otherwise, the theoretical models considered in this work 
   do not produce the wide range of line ratios along the first principal component direction 
   of the CV dataset.
In particular, 
   if we assume that the mass-transfer rate and the period are related by equation (\ref{mdot}),
   the model of LTE emission from isothermal medium  
   predicts an opposite correlation between the observed first principal component direction 
   and mass transfer rate than what is observed. 

Our analysis showed a wide variance of discriminant function values
   and a wide distribution of line ratios along the first principal component direction 
   for dwarf novae. 
In light of the known thermal cycling in the accretion disc, 
   regardless of whether the lines are emitted directly from the disc 
   or are from reprocessed radiation,  
   both the Balmer line ratios and the line equivalent widths 
   would be expected to vary with the dwarf nova outburst cycle. 
This may be a cause of the wide variance in the discriminant function values 
   and of the wide distribution along the first principal component direction. 
As far as we are aware,
   systematic observations of hydrogen Balmer emission lines throughout a dwarf nova cycle 
   have not been reported. 

Our analysis here shows that 
   the range of Bal\-mer ratios seen 
   along the first principle component direction, including ``inverted'' Balmer decrements, 
   can not be reproduced by the conventional models considered here. 
The more or less continuous variation of the line ratios 
   from Balmer decrements to Balmer ``increments'' for all CVs as a group  
   prevents any clustering of the data that clearly points 
   to a particular single underlying physical process. 
We therefore conclude that 
   either the emission from different CV sub-classes have different locations (and hence line formation processes) 
   within the system, 
   or  there are multiple regions for the hydrogen Balmer lines within a CV. 
The good matching between the data of T CrB systems 
   and the photo-ionization nebular model 
   supports the former. 
However, the latter is also likely, 
  given that the UV radiation from the accreting white dwarf 
  can ionize the cooler material in the outer accretion disc, the accretion stream 
  and the companion star.  

\section*{Acknowledgments}

GES is supported by a discovery grant 
   from the Natural Sciences and Engineering Research Council of Can\-a\-da (NSERC). 
Conversations with Laszlo Kiss about T CrB spectra are gratefully acknowledged.
KW thanks TIARA for their hospitality during his visit there 
  and for support through a Visiting Fellowship. TIARA is operated under Academia Sinica and the National
Science Council Excellence Projects programs in Taiwan
administrated through grant number NSC 94-2752-M-007-001.

\label{lastpage}

\end{document}